\documentstyle[11pt]{article}

\begin{document}

\newtheorem{theorem}{Theorem}[section]
\newtheorem{lemma}[theorem]{Lemma}
\newtheorem{defin}[theorem]{Definition}
\newtheorem{rem}[theorem]{Remark}
\newtheorem{cor}[theorem]{Corollary}
\newtheorem{prop}[theorem]{Proposition}

 %
 %
 %
 %
\def\m{\mu} \def\l{\lambda} \def\e{\epsilon} \def\r{\rho}
\def\g{\gamma}
\def\be{\begin{equation}}
\def\ee{\end{equation}}
\def\bea{\begin{eqnarray}}
\def\eea{\end{eqnarray}}
\def\beas{\begin{eqnarray*}}
\def\eeas{\end{eqnarray*}}

\def\dx{\partial_x}
\def\dv{ \partial_v }

\def\N{{\rm I\kern-.1567em N}}
\def\No{\N_0}
\def\I{{\rm I\kern-.1567em I}}
\def\R{{\rm I\kern-.1567em R}}

\def\supp{\mbox{\rm supp}}

\def\n#1{\vert #1 \vert}
\def\nn#1{\Vert #1 \Vert}

\title{Static solutions of the spherically symmetric
       Vlasov-Einstein system}
\author{G.~Rein\\
        Mathematisches Institut der Universit\"at M\"unchen\\
        Theresienstr.\ 39, W8000 M\"unchen 2, Germany}
\date{}
\maketitle
\begin{abstract}
We consider the Vlasov-Einstein system
in a spherically symmetric setting and prove the existence of
static solutions which are asymptotically flat
and have finite total mass and finite extension of the matter.
Among these there are smooth, singularity-free solutions, which have
a regular center and may have isotropic or anisotropic pressure,
and also solutions, which have a Schwarzschild-singularity at the center.
The paper is an extension of previous work, where only smooth,
globally defined solutions with regular center and isotropic pressure
were considered, cf.\ \cite{RR3}
\end{abstract}
 %
 %
 %
 %
\section{Introduction}
It is well known that the only static, spherically symmetric vacuum
solutions of Einstein's field equations are the Schwarzschild solutions
which possess a spacetime singularity (or are identically flat).
In the present note we couple Einstein's equations with the
Vlasov or Liouville equation for a static, spherically symmetric distribution
function $f$ on phase space, describing an ensemble of identical particles
such as stars in a galaxy, galaxies in a galaxy cluster etc..
This results in the following
system of equations:
\[
\frac{v}{\sqrt{1+ v^2}}\cdot \dx f -
\sqrt{1+ v^2}\, \m'\, \frac{x}{r} \cdot \dv f =0,\ x,v\in \R^3,\ r:=\n{x},
\]
\beas
e^{-2\l} (2 r \l' -1) +1 &=& 8\pi r^2 \r ,\\
e^{-2\l} (2 r \m' +1) -1 &=& 8\pi r^2 p,
\eeas
where
\beas
\r(x) = \r(r) &:=& \int_{\R^3} \sqrt{1+ v^2} f(x,v)\,dv ,\\
p(x) =p(r)  &:=& \int_{\R^3}\left(\frac{x\cdot v}{r}\right)^2 f(x,v)
\frac{dv}{\sqrt{1+v^2}} .
\eeas
Here the prime denotes derivative with respect to $r$, and
spherical symmetry of $f$ means that $f(Ax,Av)=f(x,v)$
for every orthogonal matrix $A$ and $x,v\in \R^3$.
If we let
$x=r(\sin \theta \cos\phi, \sin \theta \sin \phi, \cos\theta)$ then
the spacetime metric is given by
\[
ds^2 = - e^{2\m} dt^2 + e^{2\l} dr^2 + r^2( d\theta^2 + \sin^2\theta d\phi^2).
\]
Asymptotic flatness is guaranteed by the boundary condition
\[
\lim_{r\to\infty} \l(r) = \lim_{r\to\infty}\m(r) =0 ,
\]
a regular center at $r=0$ is guaranteed by the
boundary condition
\[
\l(0)=0 .
\]
We refer to \cite{RR1} on the particular choice
of coordinates on phase space leading to the above formulation of the system.
It should be pointed out that the above equation for $f$ is only equivalent
to the Vlasov equation if $f$ is spherically symmetric.

A brief overview of the literature related to the
present investigation seems in order.
In \cite{RR1} the initial value problem
for the corresponding time dependent system is investigated and the
existence of singularity-free solutions is established for small initial
data.
In \cite{RR2} it is shown that the classical Vlasov-Poisson system is the
Newtonian limit of the Vlasov-Einstein system.
The existence of globally defined, smooth, static solutions with a
regular center, finite radius, and finite total mass
is investigated in \cite{RR3}; that investigation is restricted to solutions
with isotropic pressure.
For the Vlasov-Poisson system the existence of spherically symmetric
steady states is established in \cite{BFH}, and we briefly recall the general
approach followed there: Taking the distribution function $f$ as a function of
the local energy and the modulus of the angular momentum, which are conserved
along characteristics, the problem is reduced to solving a nonlinear
Poisson problem. For certain cases corresponding to the so-called
polytropes it can then be shown that this leads to solutions with
finite radius and finite total mass, cf.\ also \cite{BPf}.
Cylindrically symmetric, static solutions of the Vlasov-Poisson system
are investigated in \cite{BBDP}.
A system in some sense intermediate between the Newtonian
and the general-relativistic setting is the relativistic Vlasov-Poisson
system, where similar results as for the nonrelativistic version hold,
cf.\ \cite{B}.
In \cite{SWYL} it is shown that coupling Einstein's equations with a
Yang-Mills field  can lead to static, singularity-free solutions with finite
total mass.
Static solutions of general-relativistic elasticity have been studied
in \cite{KM}.
For further references, especially on the astrophysical literature,
we refer to \cite{ST}.

The present investigation proceeds as follows: In the next section
we give two conserved quantities, which correspond to the local energy
and angular momentum in the Newtonian limit, and
reduce the existence problem for static solutions of the
Vlasov-Einstein system to solving a nonlinear
system of ordinary differential equations for $\l$ and $\m$.
The fact that the distribution
function $f$ indeed has to be a function of the local energy
and angular momentum---usually referred to as Jeans' Theorem---is
rigorously established
for the Vlasov-Poisson system in \cite{BFH} but not known in the present
situation. In other words, it is not clear whether our ansatz for $f$
covers all possible spherically symmetric, static solutions.
The existence of solutions to the remaining system for $\l$ and $\m$ is
investigated in the third section.
There we dispose of the restriction that
$f$ depends on the local energy only, which was used in \cite{RR3} and lead
to solutions with necessarily isotropic pressure. Nevertheless, the arguments
used in the present, more general situation
are considerably simpler than the ones used in \cite{RR3}; a key element
in the proof that the solutions exist globally in $r$
is the Tolman-Oppenheimer-Volkov equation.
In the fourth section we show that---for an appropriate ansatz for $f$,
corresponding to the polytropes in the classical case---the solutions
obtained by our approach have regular center, finite mass,
and finite radius.
This is done by treating the
Vlasov-Einstein system as a perturbation of the Vlasov-Poisson system
and using the criteria for a finite radius from the classical
Lane-Emden-Fowler equation.
In the last section we prove the existence of solutions with a
Schwarzschild-singularity at the center, surrounded by Vlasov-matter
with finite radius and finite total mass. Surprisingly, the latter
features of these solutions can be obtained easily and without the
perturbation argument used in the nonsingular case mentioned above.

 %
 %
 %
 %
\setcounter{equation}{0}

\section{Conserved quantities and reduction of the problem}

As explained in the introduction, a key point in our investigation is to reduce
the full system to a nonlinear system of ordinary differential equations
for $\l$ and $\m$. This is achieved by the ansatz that
the distribution function depends only on certain
integrals of the characteristic system. In the coordinates used above,
the characteristic system corresponding to the above equation for $f$ reads
\beas
\dot x &=& \frac{v}{\sqrt{1+ v^2}}, \\
\dot v &=& -  \sqrt{1+ v^2}\, \m'(r)\, \frac{x}{r} ;
\eeas
note that these equations are not the geodesic equations.
One immediately checks that the quantities
\[
E:= e^{\m(r)} \sqrt{1+v^2},\
F:= x^2 v^2 - (x\cdot v)^2 =\n{x\times v}^2
\]
are conserved along characteristics; in Sect.\ 4 the relation of $E$
to the classical local energy will become apparent,
$F$ can be interpreted as the modulus of the angular momentum.
If we take $f$ to be of the form
\[
f(x,v)= \Phi(E,F)
\]
for some function $\Phi$, the Vlasov equation is automatically satisfied.
Inserting this into the definitions of $\r$ and $p$ we obtain
the following nonlinear system for $\l$ and $\m$:
\bea
e^{-2\l} (2 r \l' -1) +1 &=& 8\pi r^2 G_\Phi (r,\m) ,\label{gll}\\
e^{-2\l} (2 r \m' +1) -1 &=& 8\pi r^2 H_\Phi (r,\m). \label{glm}
\eea
Here
\beas
G_\Phi (r,u)&=& \frac{2\pi}{r^2} \int_1^\infty \int_0^{r^2(\e^2 -1)}
\Phi(e^{\m(r)} \e, F) \frac{\e^2}{\sqrt{\e^2 -1 -F/r^2}} dF\, d\e ,\\
H_\Phi (r,u)&=& \frac{2\pi}{r^2} \int_1^\infty \int_0^{r^2(\e^2 -1)}
\Phi(e^{\m(r)} \e, F) \sqrt{\e^2 -1 -F/r^2} dF\, d\e
\eeas
for $r, u>0$
which follows by a simple transformation of variables in the integrals
defining $\r$ and $p$. These latter quantities are then given by
\[
\r(r)= G_\Phi (r,\m(r)),\ p(r)= H_\Phi (r,\m(r)),
\]
once a solution of the system (\ref{gll}), (\ref{glm}) is known to
exist.

Of course one has to make some assumptions on $\Phi$ in order to
obtain existence results for the above system and to investigate
the properties of its solutions. The assumption used below is certainly
not the weakest possible, but on the one hand it is sufficiently
general to encompass a large class of examples, leading to
quite different classes of solutions, and on the other hand the main
ideas of the proofs are not buried under technicalities.
We require $\Phi$ to satisfy the following

\bigskip
\noindent
{\bf Assumption on $\Phi$:}
\[
\Phi(E,F)= \phi (E) (F-F_0)_+^l,\ E>0,\ F>0,
\]
where $F_0\geq 0$, $l>-1/2$, and
$\phi \in L^\infty (]0,\infty[)$ is nonnegative with $\phi (E)=0,\
E>E_0$, for some $E_0 >0$.
\bigskip

Under this assumption the functions $G_\Phi$ and $H_\Phi$ are
easily seen to take the form
\bea
G_\Phi (r,u) &=& c_l r^{2l} e^{-(2l+4)u}
g_\phi \left(e^u \sqrt{1+ F_0/ r^2}\right), \label{gpdef} \\
H_\Phi (r,u) &=& \frac{c_l}{2l+3} r^{2l} e^{-(2l+4)u}
h_\phi \left(e^u \sqrt{1+ F_0/ r^2}\right), \label{hpdef}
\eea
where
\bea
g_\phi (t):=
\int_t^\infty \phi(E) E^2 (E^2 - t^2)^{l+\frac{1}{2}} dE ,\label{gdef}\\
h_\phi (t):=
\int_t^\infty \phi(E) (E^2 - t^2)^{l+\frac{3}{2}} dE \label{hdef}
\eea
for $t>0$, and
\[
c_l:= 2 \pi \int_0^1 \frac{s^l}{\sqrt{1-s}} ds .
\]
 %
 %
 %
 %

\setcounter{equation}{0}

\section{Existence of solutions}

As a first step we show that the functions $g_\phi$ and $h_\phi$
defined by (\ref{gdef}) and (\ref{hdef})
are well behaved for functions $\phi$ as considered in the assumption above.
More precisely:

\begin{lemma} \label{reggh}
Eqns.\ (\ref{gdef}) and (\ref{hdef}) define decreasing functions
$g_\phi, h_\phi \in C^1(\R^+)$ and
\[
h_\phi'(t)= - (2l+3) t \int_t^\infty \phi (E) (E^2 - t^2)^{l+1/2} dE,\ t>0.
\]
\end{lemma}

{\em Proof:}\
Obviously, the integrals defining $g_\phi$ and $h_\phi$ exist.
Lebesgue's dominated convergence theorem implies that the
functions $g_\phi$ and $h_\phi$ are continuous.
For $t>0$ and $\Delta t >0$ such that $t - \Delta t >0$ we have
\beas
&&\frac{1}{\Delta t} \left( h_\phi (t- \Delta t\right) -  h_\phi (t))
=
\frac{1}{\Delta t} \int_{t-\Delta t}^t
\phi(E) \Bigl(E^2 -(t-\Delta t)^2\Bigr)^{l+ 3/2} dE \\
&&\qquad \qquad +  \int_t^\infty
\phi(E) \frac{1}{\Delta t}\Bigl((E^2 -(t- \Delta t)^2)^{l+ 3/2}
- (E^2 - t^2)^{l+ 3/2} \Bigr)dE \\
&&\qquad\qquad =: I_1 + I_2 .
\eeas
Obviously, $I_1 \to 0$ as $\Delta t \to 0$. The term $I_2$ has a limit
as $\Delta t \to 0$ by Lebesgue's theorem. Thus, the function $h_\phi$ is
left-differentiable with
\[
\frac{d}{dt}^-  h_\phi (t) = - (2l + 3) t \int_t^\infty
\phi (E) (E^2 -t^2)^{l+ 1/2} dE .
\]
Again by Lebesgue's theorem this function is continuous, and a continuous
and continuously left-differentiable function is continuously differentiable.
Similarly, we see that
\[
\frac{d}{dt}^-  g_\phi (t) = - (2l+1) t \int_t^\infty
\phi (E) E^2 (E^2 -t^2)^{l-1/2} dE ,
\]
which is again continuous. The derivatives are negative, and thus $g_\phi$
and $h_\phi$ decrease. $\Box$

The regularity of the functions $g_\phi$ and $h_\phi$ being established,
we can now prove local existence and uniqueness of $\l$ and $\m$.
Note that in the last section we will need to pose initial data at
some $r_0>0$ which is why we include this case in the following results.

\begin{theorem} \label{lex}
Let $\Phi$ satisfy the general assumption and let $G_\Phi$ and
$H_\Phi$
be defined by Eqns.\ (\ref{gpdef}), (\ref{hpdef}), (\ref{gdef}), (\ref{hdef}).
Then for every $r_0 \geq 0$ and $\l_0, \m_0 \in \R$ with $\l_0=0$ if $r_0=0$
there exists a unique maximal solution
$\l ,\m \in C^1([r_0,R[),\ R >r_0$, of the Eqns.\ (\ref{gll}) and
(\ref{glm}) with
\[
\l(r_0)=\l_0,\ \m(r_0)= \m_0 .
\]
\end{theorem}

{\em Proof:}
Due to the regularity of the right hand sides in (\ref{gll}),
(\ref{glm})---cf.\ Lemma \ref{reggh}---the result is of course trivial in the
case $r_0>0$.
In the case $r_0=0$ one has to deal with
the singularity of the Eqns.\ (\ref{gll}) and (\ref{glm}) at $r=0$.
Using the boundary condition at $r=0$, Eqn.\ (\ref{gll}) can be
integrated to give
\be \label{darl}
e^{-2\l(r)} = 1- \frac{8\pi}{r} \int_0^r s^2 G_\Phi (s,\m(s)) ds.
\ee
If we insert this into Eqn.\ (\ref{glm}) we obtain an equation for
$\m$ alone:
\be \label{glmm}
\m'(r) =
\frac{4\pi}{1- \frac{8\pi}{r} \int_0^r s^2 G_\Phi (s,\m(s)) ds}
\left( r H_\Phi (r,\m(r)) + \frac{1}{r^2}
\int_0^r s^2 G_\Phi (s,\m(s)) ds \right) .
\ee
Integrating Eqn.\ (\ref{glmm})
subject to the initial condition $\m(0)=\m_0$ we obtain
the following  fixed point problem for $\m$:
\[
\m (r) = (T\m)(r), \ r\geq 0
\]
where
\beas
(T\m)(r):= \m_0 &+& \int_0^r
\frac{4\pi}{1- \frac{8\pi}{s} \int_0^s \sigma^2
G_\Phi (\sigma,\m(\sigma)) d\sigma} \\
&& \qquad \quad \left( s H_\Phi (s,\m(s)) + \frac{1}{s^2}
\int_0^s \sigma^2 G_\Phi (\sigma, \m(\sigma)) d\sigma \right)\,ds .
\eeas
A lengthy but straight forward argument shows that $T$ maps the set
\beas
M:= \Bigl\{ \m :[0,\delta] \to \R &\mid&
\m(0)=\m_0,\ \m_0 \leq \m(r) \leq \m_0 + 1,\\
&&\frac{8\pi}{r} \int_0^r s^2 G_\Phi (s,\m(s)) ds \leq \frac{1}{2},\
r\in[0,\delta] \Bigr\}
\eeas
into itself and acts as a contraction
with respect to the norm $\nn{\cdot}_\infty$,
provided $\delta >0$ is chosen small enough.
This gives the existence of a solution of Eqn.\ (\ref{glmm}) on the interval
$[0,\delta]$. On this interval we define
$\l$ by Eqn.\ (\ref{darl}) and obtain a local solution
$\l,\m \in C^1([0,\delta])$ of (\ref{gll}), (\ref{glm}).
Obviously, the boundary conditions at $r=0$ are satisfied,
and the solution is unique.
$\Box$

In order to show that the above solutions actually extend to
$r=\infty$ we need a relation which is known
as the Tolman-Oppenheimer-Volkov equation
in the context of the general relativistic description of spherically
symmetric fluid balls, cf.\ \cite[Eqn.\ 6.2.19]{W}.

\begin{lemma} \label{tov}
Define
\[
p_T (r):= \frac{1}{2} \int_{\R^3} \left| \frac{x \times v}{r} \right|^2
f(x,v) \frac{dv}{\sqrt{1+v^2}},
\]
and let $\l,\m$ be a solution of the system (\ref{gll}), (\ref{glm})
on the interval $[r_0,R[$. Then $p_T$ has the form
\[
p_T (r) = (l+1) p(r) + \frac{c_l}{2} F_0 r^{2l-2} e^{-(2l + 2)\mu(r)}
k_\phi\left( e^{\mu(r)} \sqrt{1+F_0/r^2}\right),
\]
where
\[
k_\phi (t) := \int_t^\infty \phi(E) (E^2 - t^2)^{l+1/2} dE,\ t>0,
\]
defines a decreasing $C^1$-function, and
\[
p'(r)= - \m'(r) \left( p(r) + \rho (r)\right) - \frac{2}{r}\left( p(r) -
p_T(r)\right),\ r\in ]r_0,R[ .
\]
\end{lemma}

{\em Proof:}
The formula for $p_T$ is obtained by a transformation of variables,
and the regularity and monotonicity of $k_\phi$ follow as in
Lemma \ref{reggh}.
Using (\ref{hpdef}), (\ref{hdef}), and Lemma {\ref{reggh} one obtains
the relation
\beas
p'(r)&=& \frac{2l}{r} p(r) - (2l+4)\,\m'(r)\, p(r) \\
&&- c_l \left(1+F_0/r^2\right) r^{2l}  e^{-(2l+2)\m(r)}\,
k_\phi \left(e^{\m(r)}\sqrt{1+F_0/r^2}\right) \m'(r) \\
&& + c_l r^{2l-3} F_0 e^{-(2l+2)\m(r)}\,
k_\phi \left(e^{\m(r)}\sqrt{1+F_0/r^2}\right)
\eeas
which, after some further calculations, leads to the Tolman-Oppenheimer-Volkov
equation. $\Box$

\bigskip
\noindent
{\bf Remark:}
The quantity $p_T$ is the tangential pressure generated by the phase space
density $f$, as opposed to the radial pressure $p$.
Note that if $F_0=0$ then $p$ and $p_T$ differ only by a multiplicative
constant, and if $F_0=l=0$, i.\ e.\ the phase space distribution is independent
of the angular momentum, then $p=p_T$, i.\ e.\ the solution has isotropic
pressure.

\begin{theorem} \label{glex}
Let $\Phi$ satisfy the general assumption and let $G_\Phi$ and
$H_\Phi$
be defined by Eqns. (\ref{gpdef}), (\ref{hpdef}), (\ref{gdef}), (\ref{hdef}).
Then for every $r_0 \geq 0$ and $\l_0 \geq 0$, $\m_0 \in \R$
with $\l_0=0$ if $r_0=0$
there exists a unique solution
$\l ,\m \in C^1([r_0,\infty[)$ of the Eqns.\ (\ref{gll}) and
(\ref{glm}) with
\[
\l(r_0)=\l_0,\ \m(r_0)= \m_0.
\]
\end{theorem}

{\em Proof:}
Let $\l, \m \in C^1([r_0,R[)$ be the maximal solution to
(\ref{gll}), (\ref{glm}) which exists by Thm.\ \ref{lex}.
We have the following equations on $[r_0,R[$:
\be
e^{-2\l} (2 r \l' -1) +1 = 8\pi r^2 \r (r),\label{l}
\ee
\be
e^{-2\l} (2 r \m' +1) -1 = 8\pi r^2 p(r), \label{m}
\ee
\be
p'(r)= - \m'(r) \left( p(r) + \r(r) \right) -
\frac{2}{r}\left( p(r) - p_T (r)\right). \label{p}
\ee
Note that---as opposed to the case considered in \cite{RR3}---the
functions $\r$ and $p$ are not necessarily strictly decreasing on their
support so that we cannot write $\r$ as a function of $p$, and we cannot
use the analysis in \cite{RS} directly.
Nevertheless, the proof given here is somewhat simpler than the one given in
\cite{RR3}.
If we integrate Eqn.\ (\ref{l}) we obtain
\be \label{ll}
e^{-2\l(r)} = 1 - \frac{8 \pi}{r} \int_{r_0}^r s^2 \r(s)\, ds
 - \frac{1}{r} r_0 (1- e^{-2\l_0}).
\ee
Inserting this into Eqn.\ (\ref{m}) yields
\be \label{mstr}
\m'(r)= 4 \pi e^{2\l(r)} \left( p(r) + w(r)\right)
\ee
where
\be \label{wdef}
w(r):= \frac{1}{r^3}
\left( \int_{r_0}^r s^2 \r(s)\, ds + \frac{r_0(1-e^{-2\l_0})}{8 \pi} \right).
\ee
Finally, by adding Eqns.\ (\ref{l}) and (\ref{m}) we have
\be \label{lmstr}
(\l'+\m')(r) = 4 \pi re^{2 \l(r)} \left( p(r) + \r(r)\right).
\ee
We now wish to establish a differential inequality for
the quantity $e^{\l + \m} (p + w)$, which will allow us to bound $\l$ and $\m$
and conclude that $R=\infty$.
Using Eqns.\ (\ref{lmstr}), (\ref{p}), (\ref{mstr}), and (\ref{wdef})
we see that
\beas
\left( e^{\l + \m}(p+w) \right)'(r)&=& e^{\l + \m} \left(
-\frac{2 p}{r}  + \frac{2 p_T}{r} - \frac{3 w}{r} + \frac{\r}{r} \right)\\
&\leq&  e^{\l + \m} \left(\frac{2 p_T}{r} + \frac{\r}{r} \right);
\eeas
note that the terms which were dropped are indeed negative by the assumption
$\l_0 \geq 0$. Now assume that $R< \infty$. Without loss of generality,
we may assume that there exists $r_1 \in ]r_0 R[$ and $C >0$ such that
$w(r) \geq C$ for $r \in [r_1,R[$, since otherwise the solution is trivial.
The fact that $\m$ is increasing and the functions $g_\phi$ and $k_\phi$
are decreasing implies that $\r$ and $p_T$ and thus also $\r/r$ and
$p_T/r$ are bounded on the interval $[r_1, R[$.
Thus, we can continue the above estimate to obtain
\[
\left( e^{\l + \m}(p+w) \right)'(r) \leq  C e^{\l + \m}
\leq C e^{\l + \m} \left(p(r) + w(r) \right).
\]
This implies that
\[
e^{\l(r) + \m(r)} \leq \frac{C}{p(r) + w(r)} \leq C,\ r\in [r_1,R[.
\]
Since $\l(r) \geq 0$ by (\ref{ll}) and $\m(r) \geq \m_0$ by monotonicity,
this implies that $\l$ and $\m$ are bounded on $[r_0,R[$ which is a
contradiction to the maximality of $R$.
Thus $R=\infty$, and the proof is complete. $\Box$

\bigskip
\noindent
{\bf Remarks:}
\begin{enumerate}
\item
If $r_0>0$ then $\l,\m \in C^2([r_0,\infty[)$ and
$\rho, p \in C^1([r_0,\infty[)$.
\item
If $r_0=F_0=0$ then
$\l,\m \in C^1([0,\infty[)\cap C^2(]0,\infty[)$ with $\l'(0)=\m'(0)=0$ and
$\rho, p \in C^1(]0,\infty[)$. If in addition $l=0$ or $l > 1/2$ then
$\l,\m \in C^2([0,\infty[) \cap C^2 (\R^3)$ and
$\rho, p \in C^1([0,\infty[) \cap C^1(\R^3)$ with $\rho'(0)=p'(0)=0$
where functions in $r$ are identified with the corresponding radially
symmetric functions on $\R^3$. Thus we see that in the case $r_0=0$
the solutions are as regular at the center as anywhere else if
various parameters are chosen appropriately.
\item
The phase space density $f:= \Phi (E,F)$ is a solution of the Vlasov equation
in the sense that it is constant along characteristics.
If $\Phi$ is continuous or continuously differentiable
then in addition $f$ has the same regularity, and in the latter case
satisfies Vlasov's equation classically.
\end{enumerate}

 %
 %
 %
 %
 %
\setcounter{equation}{0}

\section{Singularity-free solutions with finite mass and finite radius}

In this section we are interested in smooth, singularity-free solutions
with with a regular center. Thus, we consider the boundary condition
$\l(0)=0$. Also, we set $F_0=0$ in this section, the case $F_0>0$
will play its role in the next section.

In order to decide whether a solution obtained in the previous
section has finite total mass or whether $\r(r)$ vanishes
for $r$ large, one has to have rather detailed information on
the behaviour of the function $\m$. Due to the complexity of
Eqn.\ (\ref{glmm}) we have not been able to decide these questions
directly, even for simple examples of $\Phi$ and without dependence
on $F$. However, it is
possible to show that the solutions obtained above converge
to solutions of the corresponding Newtonian problem as the speed
of light tends to infinity. It is then possible to use the results
on finite mass and finite radius which are known in the Newtonian
case for the so-called polytropes to obtain solutions with
the same properties for the Vlasov-Einstein system.

To carry out this program we introduce the parameter $\g:=\frac{1}{c^2}$
where $c$ denotes the speed of light, define $\nu:= \frac{1}{\g}\m$,
and recall from \cite{RR2} that the Vlasov-Einstein system with $\g$
inserted in the appropriate places reads
\[
\frac{v}{\sqrt{1+\g v^2}}\cdot \dx f -
\sqrt{1+\g v^2} \nu' \frac{x}{r} \cdot \dv f =0,\
\]
\beas
e^{-2\l} (2 r \l' -1) +1 &=& 8\pi \g r^2 \r ,\\
e^{-2\l} (2 r \nu' +1/\g) -1/\g &=& 8\pi \g r^2 p,
\eeas
where
\beas
\r(x) &:=& \int_{\R^3} \sqrt{1+\g v^2} f(x,v)\,dv ,\\
p(x) &:=& \int_{\R^3} \left(\frac{x\cdot v}{r}\right)^2 f(x,v)
\frac{dv}{\sqrt{1+\g v^2}} .
\eeas
The conserved quantity $E$ now becomes
\[
\sqrt{1+\g v^2}e^{\g \nu(x)} ,
\]
and $F$ remains unchanged.
In order to obtain the correct limit as $\g \to 0$, we have to rewrite our
ansatz for the distribution function in the following form:
\be \label{fdef}
f(x,v) = \phi \left(\frac{1}{\g} \sqrt{1+\g v^2}e^{\g \nu(r)} - \frac{1}{\g}
\right) F^l.
\ee
As above the Vlasov-Einstein system can then be reduced to a single
equation for $\nu$ alone, namely
\bea \label{ngl}
&&\nu'(r)=\\
&&\frac{4\pi}{1- \frac{8\pi}{r} \g \int_0^r s^{2l+2} g_{\phi,\g} (\nu(s)) ds}
\left(\g r^{2l+1} h_{\phi,\g} (\nu(r)) + \frac{1}{r^2}
\int_0^r s^{2l+2} g_{\phi,\g} (\nu(s))\, ds \right) ,\nonumber
\eea
where
\be  \label{ggdef}
g_{\phi,\g} (u):=
c_l e^{-2(l+2) \g u} \int_{\frac{e^{\g u}-1}{\g}}^\infty
\phi(E) (1+\g E)^2
\left(\frac{1}{\g} (1+\g E)^2 - \frac{1}{\g} e^{2 \g u}\right)^{l+1/2} dE ,
\ee
\be \label{hgdef}
h_{\phi,\g} (u):=
\frac{c_l}{2l+3}  e^{-2(l+2) \g u}
\int_{\frac{e^{\g u}-1}{\g}}^\infty \phi(E)
\left(\frac{1}{\g} (1+\g E)^2 - \frac{1}{\g} e^{2 \g u}\right)^{l+3/2} dE ,
\ee
so that
\be \label{rpdef}
\r(r)=r^{2l} g_{\phi,\g} (\nu(r)),\ p(r)=r^{2l} h_{\phi,\g} (\nu(r)) .
\ee
For the Newtonian case the corresponding ansatz
\[
f(x,v) = \phi \left(\frac{v^2}{2} + U(r) \right) F^l
\]
reduces the Vlasov-Poisson system to the equation
\be \label{ugl}
U'(r) = \frac{4 \pi}{r^2} \int_0^r s^{2l+2} g_0 (U(s))\, ds,
\ee
where
\[
g_0 (u):= c_l \int_u^\infty \phi (E) \left(2(E - u)\right)^{l+1/2} dE .
\]
Assume that $\phi \in L^\infty(\R)$ and $E_0=0$, and fix $\nu_0 < 0$.
Clearly, the results of the previous section apply
so that for every $\g >0$ Eqn.\ (\ref{ngl}) has a unique, nontrivial, global
solution with $\nu (0)=\nu_0$. Let $U\in C^1([0,\infty[)$ be the
global solution of (\ref{ugl})  with $U(0)=\nu_0$.
We shall prove that $\nu$ converges to $U$ as $\g \to 0$, more
precisely:

\begin{lemma} \label{conv}
For every $R>0$ there exist constants $C>0$ and $\g_0 >0$ such that for every
$\g \in ]0,\g_0]$ the solution $\nu$ of (\ref{ngl}) with $\nu (0)=\nu_0$
satisfies the estimate
\[
\n{\nu (r) - U(r) } \leq C\, \g^{\min\{l+1/2,1\}} ,\ r \in [0,R],
\]
where $U$ is the solution of (\ref{ugl}) with $U(0)=\nu_0$.
\end{lemma}

{\em Proof:}\
We have the estimate
\[
\n{\nu'(r) - U'(r)} \leq I_1 + I_2 + I_3 +I_4,
\]
where
\beas
I_1&:=&
\frac{4\pi}{1- \frac{8\pi}{r} \g \int_0^r s^{2l+2} g_{\phi,\g} (\nu(s)) ds}
 \g r^{2l+1} h_{\phi,\g} (\nu(r)) ,\\
I_2&:=&
\left| \frac{1}{1- \frac{8\pi}{r} \g \int_0^r s^{2l+2} g_{\phi,\g} (\nu(s)) ds}
-1 \right|  \frac{4 \pi}{r^2} \int_0^r s^{2s+2} g_{\phi,\g} (\nu(s))\, ds, \\
I_3&:=&
\frac{4 \pi}{r^2} \int_0^r s^{2l+2} \left| g_{\phi,\g} (\nu(s))
- g_0 (\nu(s))\right| \, ds , \\
I_4&:=&
\frac{4 \pi}{r^2} \int_0^r s^{2l+2} \left| g_0 (\nu(s))
- g_0 (U(s))\right| \, ds .
\eeas
Since for $r\geq 0$ we have
$\nu(r) \geq \nu_0$
and since $\phi$ is bounded with $\phi (E)=0$ for $E>0$,
it is easily seen that
\[
g_{\phi,\g} (\nu(r)) \leq C,\ h_{\phi,\g}(\nu (r)) \leq C
,\ r\geq 0,\ \g \in ]0,1],
\]
where the constant $C$ depends only on $\phi$ and $\nu_0$.
Thus, for $R>0$ fixed and $r\in [0,R]$ we obtain the estimate
\[
I_1 \leq \frac{1}{1-C R^{2l+2} \g_0} C R^{2+1} \g \leq C \g
\]
where $\g \in ]0,\g_0]$, and $\g_0 \in ]0,1]$ is chosen such that
$1-C R^{2l+2} \g_0 >0$. Similarly,
\[
I_2 \leq C \g
\]
for $\g \in ]0,\g_0]$.
Next we estimate the difference $\n{g_{\g,\phi} (\nu (r)) - g_0( \nu (r))}$.
We can restrict ourselves to the case $\nu (r) < 0$ since otherwise
this difference is zero. Thus,
\[
\frac{e^{\g \nu(r)}-1}{\g} \geq \nu (r),
\]
which implies that we can split the difference in the following way:
\[
\n{g_{\g,\phi} (\nu (r)) - g_0( \nu (r))} \leq C (J_1 + J_2 + J_3),
\]
where
\beas
J_1&:=&
\left|e^{-2(l+2) \g \nu}-1\right|
\int_{\frac{e^{\g \nu}-1}{\g}}^\infty \phi(E) (1+\g E)^2
\left(\frac{1}{\g} (1+\g E)^2 - \frac{1}{\g} e^{2 \g \nu}\right)^{l+1/2} dE ,\\
J_2&:=&
\int_{\nu}^{\frac{e^{\g \nu}-1}{\g}} \phi (E)
\left(2 ( E - \nu)\right)^{l+1/2} dE,\\
J_3&:=&
\int_{\frac{e^{\g \nu}-1}{\g}}^\infty \phi(E) \\
&&\qquad \left|(1+\g E)^2
\left(\frac{1}{\g} (1+\g E)^2 - \frac{1}{\g} e^{2 \g \nu}\right)^{l+ 1/2}
- \left( 2(E - \nu)\right)^{l+ 1/2}\right| dE.
\eeas
Now
\beas
J_1 &\leq& C \g \frac{e^{-2(l+2)\g \nu(r)} -1}{\g} \leq C \g,\\
J_2 &\leq& C \left( \frac{e^{\g \nu (r)}-1}{\g} - \nu (r) \right)
\leq C\g ,
\eeas
and
\beas
&&
\left|(1+\g E)^2
\left(\frac{1}{\g} (1+\g E)^2 - \frac{1}{\g} e^{2 \g \nu}\right)^{l+ 1/2}
- \left( 2(E - \nu)\right)^{l+ 1/2}\right| \\
&& \qquad
\leq C \left| (1+\g E)^2 -1 \right| +
\left|
\left(\frac{1}{\g} (1+\g E)^2 - \frac{1}{\g} e^{2 \g \nu}\right)^{l+ 1/2}
- \left( 2(E - \nu)\right)^{l+ 1/2}\right| \\
&& \qquad
\leq C \left| (1+\g E)^2 -1 \right| +
C \left|
\left(\frac{1}{\g} (1+\g E)^2 - \frac{1}{\g} e^{2 \g \nu}\right)
- \left( 2(E - \nu)\right)\right|^{l+ 1/2} \\
&& \qquad
\leq C \g + C  \left| \frac{1 + 2 \g \nu (r) - e^{2\g \nu (r)}}{\g}
+ \g E^2 \right|^{\min \{l+1/2,1\}} \\
&& \qquad
\leq C \g + C  \g^{\min \{l+1/2,1\}}.
\eeas
Finally, using the method of Lemma \ref{reggh} it is easily
seen that $g_0 \in C^1(\R)$ and $\n{g_0'(u)} \leq C$ for $u\geq \nu_0$.
Thus
\[
I_4 \leq \frac{C}{r^2} \int_0^r s^{2l + 2} \n{ \nu (s) - U(s)} ds.
\]
If we combine all the above estimates we see that
\[
\n{\nu'(r) - U'(r)} \leq C \g^{\min\{l+1/2, 1\}}
+ \frac{C}{r^2} \int_0^r s^{2l+2} \n{\nu (s) - U(s)} ds
\]
for $r \in [0,R]$ and $\g \in ]0,\g_0]$. The assertion of the lemma now
follows by the usual Gronwall argument. $\Box$

In order to exploit the above result, we have to restrict our investigation
to such cases where finiteness of the radius and total mass is known
in the Newtonian situation. Thus we consider
\be \label{phidef}
\Phi (E,F):= (-E)_+^k F^l, E \in \R,\ F>0,
\ee
where $k\geq 0$ and $l>-1/2$ are such that $k<3l+7/2$.
Then we know
from \cite[5.4]{BFH} that the Newtonian potential $U$ has a
zero, which is also the radius where the density vanishes. Since by
the assumption $\nu_0 <0$ the solution is nontrivial, $U$ has to
strictly increase so that there exists $R>0$ such that $U(R)>0$.
The above lemma then tells us that for all $\g>0$ sufficiently
small, $\nu (R)>0$ which by the definition of $\Phi$ and Eqn.\
(\ref{rpdef}) implies that $\r (r) = p(r) =0$ for $r\geq R$.
Thus we have proved the following theorem:

\begin{theorem}
Let $\Phi$ be defined by Eqn.\ (\ref{phidef})
and take $\nu_0 < 0$. Then for all $\g >0$ sufficiently small the
corresponding solution $\nu$ has a zero, the density $\r$
and radial pressure $p$ as defined by
Eqns.\ (\ref{rpdef}), (\ref{ggdef}), and
(\ref{hgdef}) have finite support, and
\[
0< 4\pi \int_0^\infty s^2 \r(s) \, ds < \infty
\]
i.e.\ the solution is nontrivial with finite total mass.
\end{theorem}

\noindent
{\bf Remarks:}
\begin{enumerate}
\item
In this section we have obtained static solutions of the spherically
symmetric Vlasov-Einstein system with finite radius and finite mass,
provided $\g >0$ is sufficiently small.
However, if $f,\l,\nu$ is such a solution
then $\g^{-3/2} f(\g^{-1/2} \cdot,\g^{-1/2} \cdot), \l(\g^{-1/2} \cdot),
\g \nu (\g^{-1/2} \cdot)$ is a solution of the system with
$\g=1$, which again has finite radius and finite total mass.
\item
If $\Phi$ is defined by Eqn.\ (\ref{phidef}) then at least
$f\in C^1(\supp f)$, and $f\in C(\R^6)$ or $f\in C^1(\R^6)$
if $k>0, l\geq 0$ or $k>1,l\geq 1$ respectively.
The Vlasov equation then holds
classically on $\supp f$ or on $\R^6$ respectively.
\item
The finiteness of the total mass implies that $\l(r) \to 0$ as $r\to 0$
and $\nu$ and $\m$ have a finite limit as $r\to \infty$. This
means that the spacetime is asymptotically flat; the fact that the limit
of $\mu$ is not zero only corresponds to a rescaling
of the time coordinate $t$.
\item
Our solutions are not only global in the coordinates which we used,
but are singularity-free in the sense that the corresponding spacetime
is timelike and null geodesically complete.
\end{enumerate}
 %
 %
 %
 %
 %
 %
\setcounter{equation}{0}
\section{Solutions with a Schwarzschild singularity at the center}

In this last section we take $\Phi$ as in Sect.\ 1, with $F_0>0$,
$E_0=1$, and $\phi (E) >0$ for $E< 1$. This implies that
\beas
G_\Phi (r,u)=H_\Phi (r,u)=0  &\ \mbox{if}\ & e^u \sqrt{1+ F_0/r^2} \geq 1, \\
G_\Phi (r,u), H_\Phi (r,u)>0 &\ \mbox{if}\ & e^u \sqrt{1+ F_0/r^2} < 1 .
\eeas
Consider the field equations (\ref{gll}), (\ref{glm}). It
is well kwown that in the vacuum case, i.\ e.\ if the right
hand sides are identically zero, the asymptotically flat solutions
are given by
\[
e^{2\mu(r)} = 1 -\frac{2 M_0}{r},\
e^{2\l(r)} = \left(1 -\frac{2 M_0}{r}\right)^{-1},\ r> 2 M_0,
\]
the so-called Schwarzschild metric; $M_0 \geq 0$. This metric is
investigated in probably every textbook on general relativity, it represents
a spacetime singularity which is hidden inside of an event horizon,
a black hole. In passing we mention that the apparent singularity at
$r=2 M_0$ is only a coordinate singularity, i.\ e.\ in other coordinates
the spacetime can be extended beyond this radius, which is also called
Schwarzschild radius, cf.\ \cite[Ch.\ 6]{W}.
In the present section we wish to construct static solutions of the
spherically symmetric Vlasov-Einstein system which have such a
black hole at the center.

Let us first see if or where the Schwarzschild metric solves
our system (\ref{gll}), (\ref{glm}). This is obviously the case
if
\[
\sqrt{1-\frac{2 M_0}{r}}\sqrt{1 + \frac{F_0}{r^2}} \geq 1,
\]
i.\ e.\ for $r \in [r_-,r_+]$, where
\[
r_\pm := \frac{F_0 \pm \sqrt{F_0^2 - 16 M_0^2 F_0}}{4 F_0} .
\]
If we take $F_0 > 16 M_0^2 >0$ we obtain $ 2 M_0 < r_- < r_+$, and we may set
\[
e^{2\mu(r)} = 1 -\frac{2 M_0}{r},\
e^{2\l(r)} = \left(1 -\frac{2 M_0}{r}\right)^{-1}
\]
and
\[
\r (r) = p(r)= f(x,v) =0
\]
for $2 M_0 < r \leq r_+$. At $r_0:= r_+$ we now pose initial conditions
\[
\mu (r_0)= \mu_0 := \sqrt{1-\frac{2 M_0}{r_0}},
\l (r_0)= \l_0 := \sqrt{1-\frac{2 M_0}{r_0}}^{-1},
\]
and solve (\ref{gll}), (\ref{glm}) to the right of $r_0$ according
to Thm.\ \ref{glex}; note that $\l_0>0$. In this way we obviously
obtain a static solution of the spherically symmetric Vlasov-Einstein
system, which coincides with the Schwarzschild solution for
$r\leq r_0$. The phase space density $f$ is no longer given by
one function of $E$ and $F$ because the vacuum solution is not
consistent with our ansatz $f=\Phi (E,F)$ in the region
$]2 M_0,r_-[$, but $f$ still solves the Vlasov equation everywhere
in the region $]2 M_0, \infty[$ since no characteristics
which carry mass can cross the region $]r_-,r_0[$.

What remains to be seen is that our solution is not vacuum everywhere
for $r> 2 M_0$, i.\ e.\ we really get Vlasov matter for $r>r_0$, and
that this matter outside $r_0$ again has finite radius and finite
total mass.
First note that
\[
\left( e^{\m(r)} \sqrt{ 1 + \frac{F_0}{r^2}} \right)'_{|r=r_0} <0
\]
which implies that $\r (r) >0$ and $p(r) >0$ in a right neighborhood of
$r_0$. To show that for large enough values of $r$ the density
vanishes again, it is obviously sufficient to show that
$\lim_{r\to \infty} \m (r) > 0$. To see the latter we first recall that
by integrating Eqn.\ (\ref{gll}) we obtain
\[
e^{-2 \l(r)} = 1 - \frac{2 m(r)}{r} - \frac{1}{r} r_0 (1- e^{-2\l_0})
= 1 - \frac{2 M(r)}{r},\ r> 2 M_0,
\]
where
\[
M(r):= M_0 + m(r) := M_0 + 4 \pi \int _{r_0}^r s^2 \r(s)\, ds .
\]
Inserting this into Eqn.\ (\ref{glm}) we obtain
\beas
\m'(r) &=& \frac{1}{1-\frac{2 M(r)}{r}}
\left(\frac{M(r)}{r^2} + 4 \pi r p(r) \right) \\
&\geq&
\frac{M_0}{r(r-2 M_0)} + \frac{m(r)}{r^2} ,
\eeas
and integrating this yields
\beas
\m(r) &\geq& \m_0 + M_0 \int_{r_0}^r \frac{ds}{s(s-2M_0)}
+ \int_{r_0}^r \frac{m(s)}{s^2} ds \\
&=& \m_0 + \ln \sqrt{1-\frac{2 M_0}{r}} - \ln \sqrt{1- \frac{2 M_0}{r_0}} +
+ \int_{r_0}^r \frac{m(s)}{s^2} ds \\
&=&   \ln \sqrt{1-\frac{2 M_0}{r}}
+ \int_{r_0}^r \frac{m(s)}{s^2} ds .
\eeas
Therefore
\[
\lim_{r\to \infty} \m(r) \geq \int_{r_0}^\infty \frac{m(s)}{s^2} ds >0,
\]
which proves that the density vanishes for $r$ large enough.
Note also that the inner boundary of the Vlasov matter satisfies the
estimate
\[
r_0 > \frac{F_0}{4 M_0} > 4 M_0 .
\]
We collect these results in the following theorem:

\begin{theorem} \label{sing}
Let $\Phi$ satisfy the general assumption in Sect.\ 1 with $E_0=1$ and
$\phi (E)>0$ for $E >1$, and let $F_0 > 16 M_0^2 >0$.
Then there exists a static solution $(f,\l,\m)$ of the spherically
symmetric Vlasov-Einstein system such that
\[
e^{2\mu(r)} = 1 -\frac{2 M_0}{r},\
e^{2\l(r)} = \left(1 -\frac{2 M_0}{r}\right)^{-1},\ 2 M_0 < r < r_0,
\]
\[
\r(r)= p(r) = f(x,v)=0,\ 2 M_0 < r < r_0 \ \mbox{or}\ r>R ,
\]
and
\[
0 < 4 \pi \int _{r_0}^R \r(r) dr < \infty ,
\]
where
\[
r_0 := \frac{F_0 + \sqrt{F_0^2 - 16 M_0^2 F_0}}{4 F_0} > 4 M_0
\]
and $R > r_0$. Furthermore,
$\l,\m \in C^2 (]2 M_0, \infty[)$, $\r, p \in C^1 (]2 M_0, \infty[)$,
and the spacetime in asymptotically flat.
\end{theorem}

\noindent
{\bf Remarks:}
\begin{enumerate}
\item
As pointed out above, the phase space density $f$ is not given as a function
of $E$ and $F$ globally for $r> 2 M_0$.
\item
If we start at the center
with a smooth static solution as obtained in Sect.\ 4 instead
of the Schwarzschild singularity we
can use the method of the present section to obtain static solutions
of the following kind:
For $0\leq r \leq r_0$ the nontrivial matter distribution is given by
$f(x,v)= \Phi_0 (E)$, for $r_0 < r < r_1$ we have vacuum, for
$r_1 \leq r \leq r_2$ the matter distribution is again nonzero and
given by $f(x,v)= \Phi_1 (E,F)$, and this procedure can be continued.
The resulting solution is smooth, geodesically complete, asymptotically
flat, with finite mass and finite radius, and consists of rings of Vlasov
matter
separated by vacuum.
Contrary to our expectation it is not clear that such a solution violates
Jeans' Theorem, which in spite of 1) might still be valid for smooth
solutions. Note however that as the radius increases,
the pressure can change from isotropic to unisotropic.
\item
In the Newtonian case one can construct analogous solutions, where the
role of the Schwarzschild singularity is played by a point mass
$M_0$ situated at the center $r=0$. It then turns out that the resulting
solution does not violate Jeans' Theorem---in the above notation,
$r_-=0$ in the Newtonian case. Thus one can at least say that the
range of validity of Jeans' Theorem is larger in the classical case
than in the general relativistic. One can also construct solutions
of the type described in 2) for the Newtonian case.
\end{enumerate}

\end{document}